\documentclass{article}

\usepackage{amsmath,amsfonts}
\usepackage{microtype}
\usepackage{graphicx}
\usepackage{subcaption}
\usepackage{booktabs} 
\usepackage{enumitem}
\usepackage[square,sort,comma,numbers]{natbib}
\setlist{nolistsep}

\usepackage{hyperref}

\usepackage[accepted]{icml2021}

\icmltitlerunning{Annotating sleep states}

\begin{document}

\twocolumn[
\icmltitle{Annotating sleep states in children from wrist-worn accelerometer data using Machine Learning}



\icmlsetsymbol{equal}{*}

\begin{icmlauthorlist}
\icmlauthor{Ashwin Ram}{equal,ece}
\icmlauthor{Sundar Sripada V. S.}{equal,ece}
\icmlauthor{Shuvam Keshari}{equal,ece}
\icmlauthor{Zizhe Jiang}{orie}
\end{icmlauthorlist}

\icmlaffiliation{ece}{Chandra Department of Electrical and Computer Engineering, UT Austin}
\icmlaffiliation{orie}{Department of Operations Research and Industrial Engineering, UT Austin}

\icmlcorrespondingauthor{Shuvam Keshari}{skeshari@utexas.edu}

\icmlkeywords{Machine Learning, Classification, Boosting, Sleep states}

\vskip 0.3in
]



\printAffiliationsAndNotice{\icmlEqualContribution} 

\begin{abstract} \label{sec:abstract}
Sleep detection and annotation are crucial for researchers to understand sleep patterns, especially in children. With modern wrist-worn watches comprising built-in accelerometers, sleep logs can be collected. However, the annotation of these logs into distinct sleep events -- \textit{onset} and \textit{wakeup}, proves to be challenging. These annotations must be automated, precise, and scalable. We propose to model the accelerometer data using different machine learning (ML) techniques such as support vectors, boosting, ensemble methods, and more complex approaches involving LSTMs and Region-based CNNs. Later, we aim to evaluate these approaches using the Event Detection Average Precision (EDAP) score (similar to the IOU metric) to eventually compare predictive power and model performance. Our GitHub repository can be viewed \href{https://github.com/ss26/ece381k-applied-ml-project}{\underline{here}}.
\end{abstract}


\section{Introduction}
\label{sec:intro}

\subsection{Why is sleep detection important?} 

Sleep is crucial in regulating mood, emotions, and behavior in individuals of all ages, particularly children \cite{vandekerckhove2018emotion}. By accurately detecting periods of sleep and wakefulness from wrist-worn accelerometer data, researchers can gain a deeper understanding of sleep patterns and better understand disturbances caused by sleep in children \cite{robbins2019sleep, de2019wearable}. 

\subsection{How does data science play a role in sleep detection?}

So far in sleep literature, the most accurate way to annotate sleep events is by using sleep logs. However, the manual effort required to maintain these logs over time engenders nuances in the annotation; a person going to bed versus actually falling asleep may be at very different times, which may fabricate errors in these logs \cite{robbins2019sleep}. 

In contrast, a wrist-worn accelerometer can track human-engineered sleep-related features (like arm angle) which can potentially better help detect and annotate sleep events. However, these features vary across individuals making it hard for sleep researchers to accurately identify sleep windows. By leveraging the right Machine Learning (ML) approaches, it is possible to learn the higher dimensional sleep data and accurately annotate sleep events, thereby bridging the gap between data availability and data analyzability \cite{van2018estimating, sundararajan2021sleep}.

\subsection{Related Work}
\label{sec:related_work}

The rapidly evolving domain of sleep pattern detection through wearable technology has gained considerable traction in recent times. A standout contribution in this area is by \cite{sano2018multimodal}, who have ingeniously employed smartphones and wearables to introduce a non-intrusive method for ambulatory sleep tracking. Leveraging the power of Long Short-Term Memory (LSTM) Recurrent Neural Networks, they have adeptly determined sleep/wake states by assimilating temporal information. Their exhaustive research, spanning 5580 days and involving 186 participants, reported an admirable sleep/wake classification accuracy of $96.5\%$. Remarkably, when pitted against traditional non-temporal machine learning techniques and well-regarded actigraphy software, the LSTM model demonstrated unparalleled efficacy, signifying the substantial potential of recurrent neural networks in sleep detection. Reflecting on the ongoing Kaggle \href{https://www.kaggle.com/competitions/child-mind-institute-detect-sleep-states/overview}{\underline{competition}}, many teams are extensively engineering features on `train' and `test' datasets, originally extracted from Parquet files. They craft temporal attributes from the 'timestamp' column, optimize memory usage, and meticulously introduce new features employing statistical techniques, including rolling metrics and lag/lead shifts. This intricate feature engineering is orchestrated in batches to manage voluminous data, followed by a correlation analysis visualized through a heatmap. Techniques like random forest, Xgboost, and light-GBM serve as the backbone for training. Drawing from \cite{sano2018multimodal}'s groundbreaking work, it's evident that LSTM models, particularly in real-time scenarios, could be the game-changer in advancing sleep detection, thus enhancing holistic health monitoring.
\section{Problem Description}
\label{sec:problem_description}

In this project, we analyze wrist-worn accelerometer data for sleep monitoring to predict two important sleep states: \textit{onset} and \textit{wakeup}. The project is a submission to the code competition hosted on Kaggle: \href{https://www.kaggle.com/competitions/child-mind-institute-detect-sleep-states/overview}{Child Mind Institute (CMI) - Detect Sleep States}. This is an ongoing competition that is set to be completed on November 28, 2023 -- which gives us (the authors) the perfect timeline to participate.

In this section, we provide the source for our data and explain the different files and the different columns of data. 

\subsection{Data Explanation}

All data required for the project is provided by the competition, at \href{https://www.kaggle.com/competitions/child-mind-institute-detect-sleep-states/data}{this link.} The data is provided by the Healthy Brain Network, a landmark mental health study based in New York City partnered with the CMI. 

The data consists of 500 multi-day recordings of accelerometer data. The two types of events are \textit{onset}, and \textit{wakeup}. For accelerometer data, sleep is referred to as the longest period of inactivity; no movement recorded in the watch means the wearer is probably sound asleep. 

We now briefly describe details about sleep periods:
\begin{itemize}
  \setlength\itemsep{0pt}
    \item A single sleep period must be at least 30 minutes in length, 
    \item A single sleep period can be interrupted by bouts of activity that do not exceed 30 consecutive minutes,
    \item No sleep windows can be detected unless the watch is deemed to be worn for the duration
    \item The longest sleep window during the night is the only one which is recorded
    \item If no valid sleep window is identifiable, neither an \textit{onset} nor a \textit{wakeup} event is recorded for that night.
\end{itemize}

Sleep events don't need to straddle the day-line, and therefore there is no hard rule defining how many may occur within a given period; however, no more than one window should be assigned per night. For example, it is valid for an individual to have a sleep window from 01h00–06h00 and 19h00–23h30 on the same calendar day, though assigned to consecutive nights. There are as many nights recorded for a series as there are 24-hour periods in that series.

For some periods in the data, the accelerometer was removed. This can be determined by suspiciously little variation in the accelerometer signals over a long period. These periods must not be annotated by our model; annotation will result in these events being scored as false positives (explained further in Section~\ref{sec:metrics}).

\begin{figure}[ht]
    \centering
     \begin{subfigure}[b]{\linewidth}
         \centering
         \includegraphics[width=\textwidth]{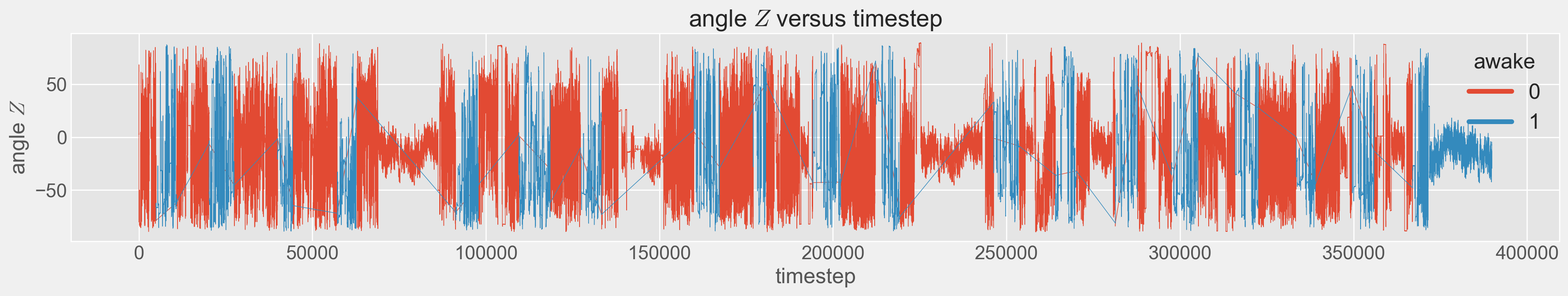}
     \end{subfigure}
     \hfill
     \begin{subfigure}[b]{\linewidth}
         \centering
         \includegraphics[width=\textwidth]{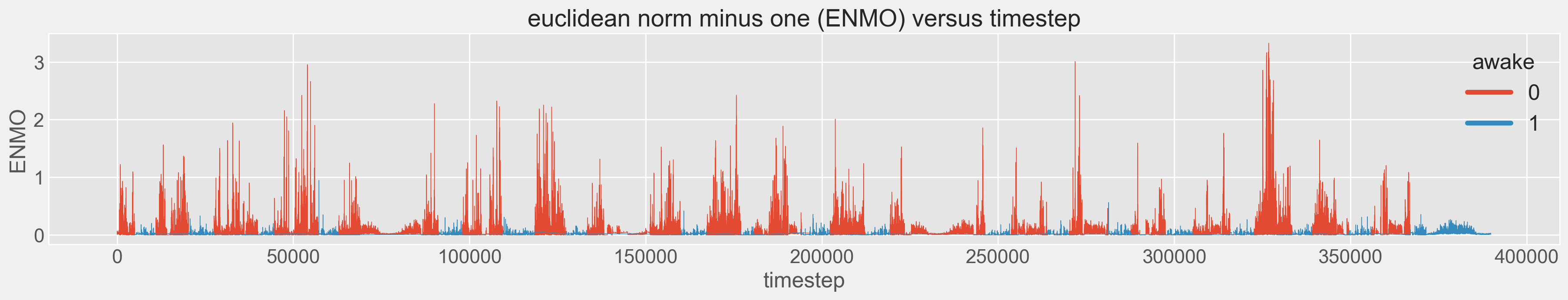}
     \end{subfigure}
    
    \caption{\small \textbf{AngleZ, ENMO versus timestep.} This plot shows angleZ and ENMO (as explained in \ref{sec:feature_columns}) to illustrate how the accelerometer data is recorded over time. The red and blue portions show the sleeping periods and the awake periods respectively.}
    \label{fig:anglez_enmo}
\end{figure}

Here, we describe the different columns in \textbf{train\_series.parquet}, which contain series to be used as training data. Each series is a continuous recording of accelerometer data for a single subject spanning many days.

\begin{enumerate}\label{sec:feature_columns}
    \item \textbf{series\_id} - unique identifier for each accelerometer series,
    \item \textbf{step} - an integer timestep for each observation within a series,
    \item timestamp - a corresponding datetime with ISO 8601 format \%Y-\%m-\%dT\%H:\%M:\%S\%z,
    \item \textbf{anglez} - as calculated and described by the \href{https://cran.r-project.org/web/packages/GGIR/vignettes/GGIR.html#4_Inspecting_the_results}{GGIR package}, z-angle is a metric derived from individual accelerometer components that are commonly used in sleep detection, and refers to the angle of the arm relative to the vertical axis of the body; figure~\ref{fig:anglez_enmo} shows a scatter plot of \textbf{anglez} as a function of the timestep, 
    \item \textbf{enmo} - as calculated and described by the \href{https://cran.r-project.org/web/packages/GGIR/vignettes/GGIR.html#4_Inspecting_the_results}{GGIR package}, ENMO is the Euclidean Norm Minus One of all accelerometer signals, with negative values rounded to zero.; while no standard measure of acceleration exists in this space, this is one of the several commonly computed features.
\end{enumerate}

Now, we describe the different columns present in \textbf{train\_events.csv}, which contains sleep logs for series in the training set recording onset and wake events.

\begin{enumerate}
    \item \textbf{series\_id} - unique identifier for each series of accelerometer data in \textbf{train\_series.parquet},
    \item \textbf{night} - an enumeration of potential onset/wakeup event pairs; at most one pair of events can occur for each night,
    \item \textbf{event} - the type of event, whether onset or wakeup,
    \item \textbf{step} and \textbf{timestamp} - the recorded time of occurrence of the event in the accelerometer series.
\end{enumerate}

The train\_series corresponds to `X\_train' and contains 128 million rows and 5 columns, with 277 unique series\_id. The train\_events correspond to `y\_train' and test\_series correspond to `X\_test' which will be evaluated later using \ref{sec:edap}

\section{Methodology}
\label{sec:methodology}
To rigorously address the task of accurately detecting the onset and wakeup events within the accelerometer data, we aim to conduct a comparative study leveraging both traditional ML models and advanced deep learning networks. In the following sections, we describe each of the considered methods and provide expository blurbs to summarize them. To begin with, we will explore the nature of the data for insights into feature engineering.

\subsection{EDA: Exploratory Data Analysis}

First, we notice the nature of the data in the zoomed-in area of angleZ vs timestep as shown below:

\begin{figure}[h]
    \centering
    \includegraphics[width=\linewidth]{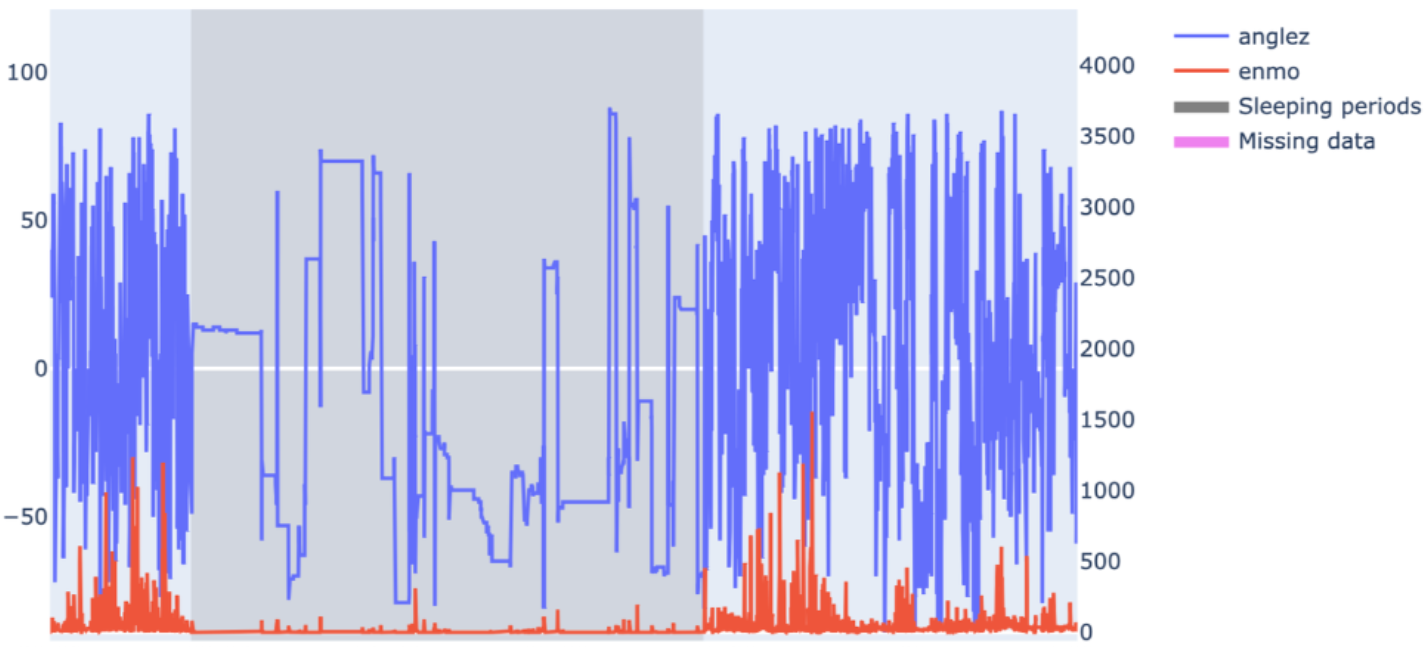}
    \caption{\small \textbf{Zoomed in angleZ vs timestep} This plot shows angleZ (as explained in \ref{sec:feature_columns}) to illustrate there are distinct low-frequency and high-frequency components to the signal during the `sleep' and `awake' states}
    \label{fig:eda}
\end{figure}

We can see that there appear to be distinct low-frequency and high-frequency components to the signal during the `sleep' and `awake' states. This gives us insights into feature engineering. For both enmo and anglez, we generate rolling mean, max, and standard deviation values for window sizes corresponding to 5 min, 30 mins, 2 hrs, and 8 hrs. We add these new features to the training dataset.

\subsection{Traditional ML Models}
For baseline comparisons, we employ traditional machine learning models, which include but are not limited to the following four broad approaches. 

\subsubsection{Support Vector Machines (SVMs)} 

Concerning high-dimensional spaces, such as those we encounter in multi-day accelerometer readings, SVMs \cite{scholkopf2000new} stand out for their classification prowess. Specifically, we will employ an SVM with a radial basis function (RBF) kernel, an imperative to adeptly maneuver the non-linearities observed in our time-series data.

\subsubsection{Random Forests (RF)} 

Random Forests \cite{breiman2001random} are integral to our approach, primarily due to their renowned accuracy in scenarios dealing with extensive datasets characterized by higher dimensionality, akin to our continuous accelerometer recordings. Notably, RF's ability to quantify feature importance stands paramount, offering valuable insights into the influential determinants underpinning model decisions.

The methodology involves using the Polars library for data manipulation, handling the Child Mind Institute's dataset, and extracting features inspired by sleep data exploration. This is followed by the import and implementation of an Event Detection AP score function, which is essential for validating the model's results before submission. A notable aspect of this code is the feature engineering process, where features such as the current hour, rolling aggregates (mean, max, std) of certain variables over different time windows, and the total variation of these variables are computed. This detailed approach to feature engineering is based on the observation that during sleep, certain physiological parameters exhibit characteristics akin to jump processes, whereas during wakefulness, they resemble diffusion processes. The total variation, a measure of the cumulative absolute differences between points in a time series, is used to differentiate these states effectively.

The training process involves splitting the dataset into training and validation sets, normalizing the features, and fitting the models. The Random Forest classifier is trained with specific parameters like the number of estimators and minimum samples per leaf. The importance of each feature in the model is analyzed using a bar plot, providing insights into which features contribute most significantly to the model's decision-making process. After training, the models are validated on a separate dataset, and their performance is assessed using the previously mentioned Event Detection AP score function. Finally, the trained model is applied to the test dataset to predict sleep states. This comprehensive approach, blending feature engineering with advanced machine learning techniques, demonstrates the potential of machine learning in enhancing the analysis and understanding of complex healthcare data, particularly in the crucial area of sleep state detection.

\subsection{Modern ML Models}

\subsubsection{Long Short-Term Memory (LSTM) Recurrent Neural Networks}

Given the sequential nature of our data, LSTMs \cite{sherstinsky2020fundamentals} are ideal for capturing long-range dependencies and temporal dynamics inherent in sleep patterns. 

Our model's architecture is carefully crafted to address the nuances of time-series data characteristic of pediatric sleep patterns. At its core, the LSTM layer with 64 units is pivotal for capturing the temporal dependencies in the sleep data. This choice is motivated by the LSTM's proven capability to handle long sequences, which is essential for monitoring sleep over extended periods. The inclusion of a dropout layer with a 0.7 rate following the LSTM layer is a strategic decision to combat overfitting, a common challenge in deep learning models. This layer randomly deactivates a fraction of the input units, thus ensuring that the model does not rely on any specific set of features too heavily. The architecture is finalized with a dense layer featuring a sigmoid activation function. This layer's role is to consolidate the learned features into a binary output, signifying the sleep state.

The preprocessing of our data is a multi-faceted process, crucial for the LSTM model's performance. We initiate this process by importing data from Parquet files, a format chosen for its efficiency in handling large datasets. The data, comprising various sensor readings, is timestamped, necessitating a conversion into a sequential numerical format suitable for time series analysis. We employ Pandas and PyArrow libraries for these tasks, capitalizing on their robust data manipulation capabilities. The transformation stage involves normalizing the data, an essential step to standardize the input range for the LSTM model. Furthermore, we meticulously handle missing values, a critical step to maintain the chronological integrity of the time series data. 

The training of our LSTM model is an elaborate process. We begin by preparing separate datasets for training and validation, ensuring a proper split that reflects the model's performance in unseen data scenarios. The model is compiled with the Adam optimizer, chosen for its effectiveness in handling sparse gradients and its adaptive learning rate capabilities, and binary cross-entropy as the loss function, suitable for our binary classification task. To enhance the model's training, we implement callbacks such as ModelCheckpoint, for saving the model at optimal states, and EarlyStopping, to prevent overfitting by halting the training when validation loss ceases to decrease. Additionally, we address the potential issue of class imbalance in our dataset by employing a custom class weighting strategy, thereby ensuring that our model is equally sensitive to all sleep states. The training is executed over a predefined number of epochs, with careful monitoring of both training and validation accuracy and loss, allowing us to gauge the model's learning progression and generalization capabilities. 

\subsubsection{Region-Based Convolutional Neural Networks (RCNNs)}

Given the dataset comprised time-stamped images annotated with sleep state labels, the initial preprocessing included the installation of pycocotools and loading of necessary Python libraries. A custom data handler, SleepStatesDataset, was implemented for efficient data feeding into the model. The dataset was partitioned into training (70 percent), validation (20 percent), and test (10 percent) sets, ensuring a balanced representation of classes.

The model architecture was based on a Faster RCNN with a ResNet50 backbone, leveraging transfer learning. The classifier head of the pre-trained model was modified to adapt to our specific three-class problem (two sleep states plus background). The training was conducted over several epochs with a specified learning rate and Adam optimizer. The model was evaluated on the validation set at the end of each epoch, focusing on average precision as the primary metric. Custom functions were implemented to streamline the training and evaluation process, ensuring computational efficiency.

The performance evaluation relied on the computation of the mean Average Precision (mAP) over the validation and test sets. This metric was chosen due to its relevance in object detection tasks, particularly in scenarios with imbalanced classes.

The RCNN model demonstrated promising results in detecting and classifying sleep states. During training, the model exhibited a consistent decrease in loss metrics, indicating effective learning. On the validation set, the model achieved an mAP of 0.76, reflecting its robustness in handling diverse data. Visual inspection of the model's predictions on test images further corroborated its practical efficacy.

\section{Metrics \& Evaluation}
\label{sec:metrics}


We first define the terms used in the evaluation. We refer to each \textit{onset} and \textit{wakeup} entry as an \texttt{event}. We refer to timestamp error tolerance as \texttt{tolerance}. 

Now, we explain the scoring scheme. For each \texttt{event} $\times$ \texttt{tolerance} group, we calculate the Average Precision (AP) score. The average precision score is the area under the precision-recall curve generated by decreasing confidence score thresholds over the predictions. 

In this calculation, matched predictions over the threshold are scored as TP (true positives) and unmatched predictions as FP (false positives). Unmatched ground-truths are scored as FN (false negatives).

Due to the nature of the Kaggle competition, the scoring is handled by the competition's submission software. 

\subsection{EDAP: Event Detection Average Precision}
\label{sec:edap}
EDAP is an AUC-PR metric for event detection in time series and video.

Similar to IOU-threshold average precision metrics commonly used in object detection. For events occurring in time series, we replace the IOU threshold with a time tolerance.

Evaluated on the average precision of detected events, averaged over timestamp error tolerance thresholds, and averaged over event classes.

\begin{itemize}
    \item \textbf{Selection} - (optional) Predictions not within a series' scoring intervals are dropped.
    \item \textbf{Assignment} - Predicted events are matched with ground-truth events.
    \item \textbf{Scoring} - Each group of predictions is scored against its corresponding group of ground-truth events via Average Precision.
    \item \textbf{Reduction} - The multiple AP scores are averaged to produce a single overall score.
\end{itemize}

\section{Results} \label{sec:results}

\subsection{SVM with RBF kernel}
We used the Support Vector Classification with Radial Basis Function (RBF) kernel. We prototyped using 2 million rows out of 128 million.
Only the ‘2-hr’ features were used (6 out of 25 features). We found that it was computation-intensive on Google Colab. The EDAP score we obtained was 0.23, likely due to training on a very small portion of data. This study was done to understand how the evaluation metric works and was a prototype model to get us started with exploring better options.

\subsection{Random Forest}
The Random Forest model's feature importance graph reveals key predictors in the sleep state detection task, with features derived from 'enmo' and 'anglez' variables—particularly their rolling statistics over short periods—dominating in importance. The prominence of $'anglez\_30m\_max'$ and $'enmo\_120m\_mean'$ underscores the model's sensitivity to variations in movement and arm angle, which are indicative of sleep-related inactivity. Lesser importance assigned to features like 'hour' suggests that time of day may be less critical than physical stillness or movement patterns in predicting sleep states.

\begin{figure}[h]
    \centering
    \includegraphics[width=\linewidth]{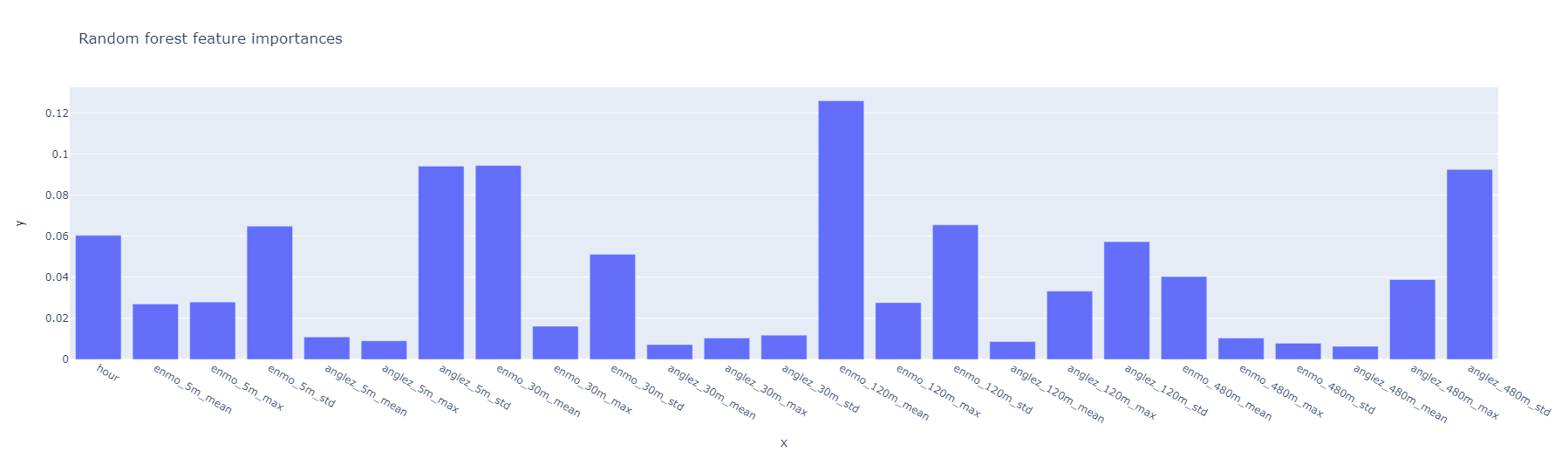}
    \caption{\small \textbf{RF feature importances} This plot shows the feature importances. 2-hr features seem to be the most important (probably because sleep cycles are 90-120 mins long?}
    \label{fig:rf}
\end{figure}

\subsection{LSTM Results}
In the initial training, the LSTM model's validation loss decreased significantly from infinity to 0.28741, and training accuracy began at 65.30\%. Over epochs, training accuracy improved to 73.06\%, but validation loss increased slightly, indicating generalization challenges. Validation accuracy peaked at 90.28\%, suggesting initial overfitting, later mitigated. The model demonstrated learning in classifying sleep states but started to plateau, implying a need for model adjustments. Training was efficient, with total CPU and wall times under two minutes.

Below is a visualization of the model's prediction performance over time:

\begin{figure} [h]
\centering
\includegraphics[width=\linewidth]{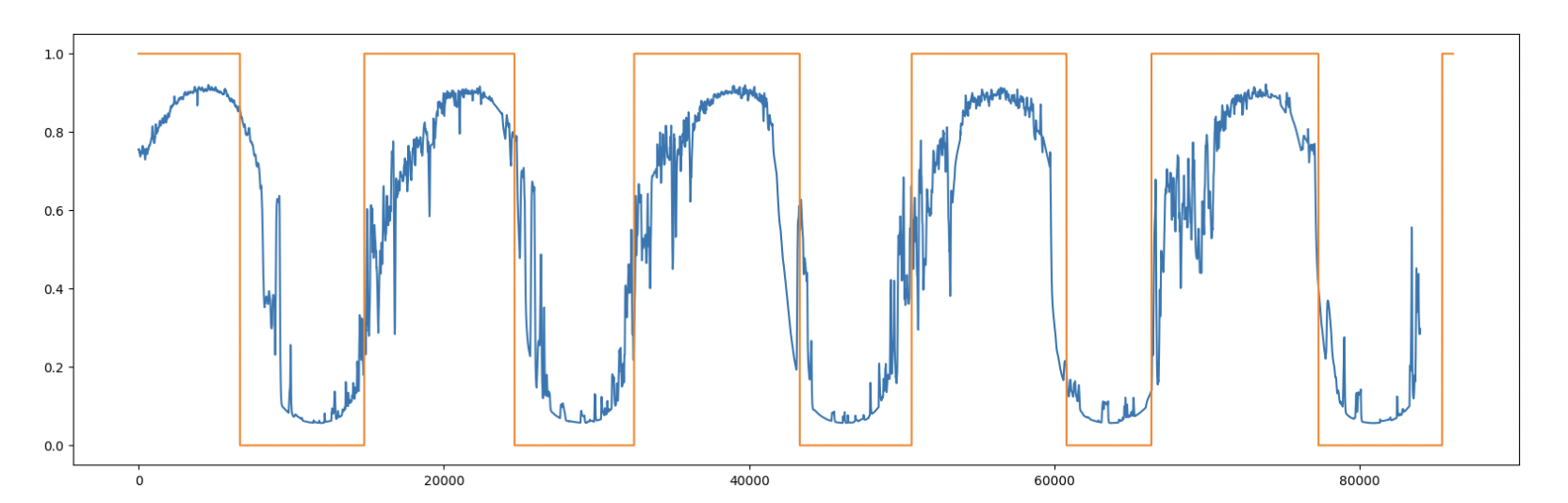}
\caption{The LSTM model's predictions are plotted against the ground truth for sleep state detection. The model's predictions closely follow the cyclical pattern of the ground truth, indicating effective learning from the time-series data.}
\label{fig:sleep_state_prediction}
\end{figure}

\subsection{RCNN Results}
In our innovative approach, we apply Region-based Convolutional Neural Networks (RCNN) to the domain of child sleep state detection using time series data from wrist-worn accelerometers. This intersection of time series analysis and object detection methodologies presents a novel technique in pediatric sleep research.

The accelerometer data, represented as a time series, is transformed into a format suitable for object detection. Each data point in the time series is analogous to a pixel in an image. This conceptual transformation allows us to apply RCNN, a technique traditionally used in image processing, to our time series data. The data is segmented into windows, each window acting as an `image' for the RCNN model.

Annotations for sleep states are akin to bounding boxes in traditional object detection tasks. The central coordinate $x$ of a bounding box in our scenario corresponds to the midpoint of a time interval, representing a specific sleep state.

\begin{equation}
x = \frac{x_0 + x_1}{2}
\end{equation}

where $x_0$ and $x_1$ are the start and end of the annotated sleep state interval.

The RCNN model processes these transformed data windows to predict sleep states. The predictions include the start and end of a sleep state, the label of the sleep state, and the confidence score. We apply a threshold $\tau$ to filter predictions:

\begin{equation}
B = \{ b | s(b) > \tau \}
\end{equation}

where $s(b)$ is the confidence score of a predicted sleep state interval $b$.

The model's output is visualized by plotting the predicted sleep states over the transformed time series data. The predicted intervals are marked in cyan, while the ground truth annotations are represented by yellow lines. This visual representation allows for an intuitive comparison between the model's predictions and the actual sleep states.

\begin{figure} [h]
\centering
\includegraphics[width=\linewidth]{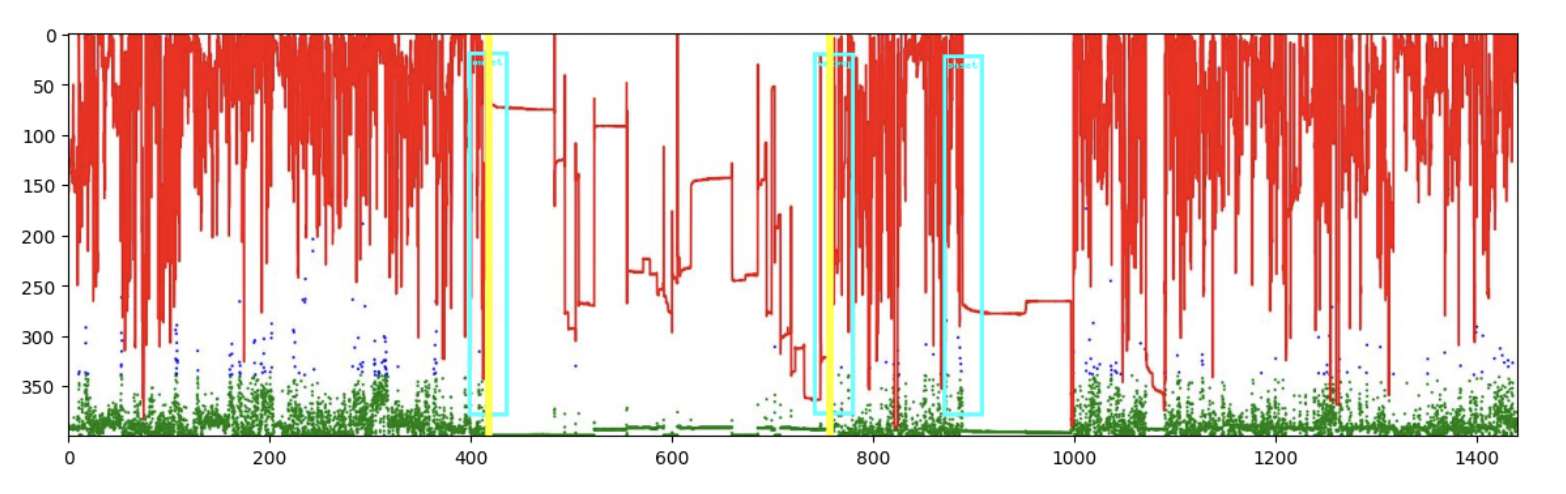}
\caption{RCNN, plotting the predicted sleep states over the transformed time series data.}
\label{fig:rcnn}
\end{figure}

\begin{figure}[h]
\centering
\includegraphics[width=0.8\linewidth]{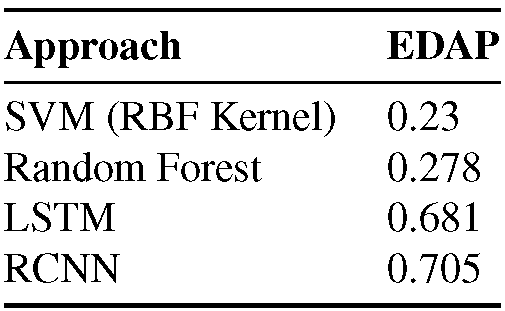}
\caption{Our list of approaches and their Event Detection Average Precision scores}
\label{fig:tab}
\end{figure}

Our results demonstrate the potential of using RCNN models in interpreting time series data from accelerometers for child sleep state detection. The approach of treating time intervals as bounding boxes provides a unique perspective in analyzing time series data, especially in the context of pediatric sleep patterns. All the results are summarized in Figure ~\ref{fig:tab}

\section{Limitations and Future Work}
In our project on using machine learning to annotate children's sleep states from wrist-worn accelerometer data, we encountered several limitations and identified areas for future work. The specificity of our dataset to wrist-worn devices may limit the generalizability of our findings to other types of sleep monitoring tools. Advanced models like LSTM and RCNN, while effective, pose challenges due to their computational intensity, which might restrict their practical application in larger or real-time settings. Additionally, the variability in children's sleep patterns presents a hurdle for model accuracy and generalization. Looking ahead, we see the potential for expanding our research to include cross-device validation and further optimization of our models to balance computational efficiency with accuracy. Investigating personalized models that can adapt to individual variability in sleep patterns could also enhance the accuracy of sleep state annotations.
\section{Conclusion}
This study contributes significantly to the field of pediatric sleep monitoring by applying various machine learning techniques to wrist-worn accelerometer data. While there are challenges related to data specificity, model complexity, and sleep pattern variability, the research provides a foundation for future advancements in automated sleep state annotation, with potential benefits for clinical and research applications in pediatric sleep studies.

In terms of lessons learned, we all gained insights into the complexities of dealing with real-world data, particularly in the context of pediatric health and wearable technology. The process of selecting and fine-tuning machine learning models to suit specific tasks taught us the importance of methodical approach and critical evaluation. Balancing computational efficiency with the EDAP metric, especially in advanced models like LSTM and RCNN, was a key lesson. Overall, the project not only advanced our understanding of pediatric sleep monitoring but also provided us with solid practical experience in machine learning applications.

\bibliography{main}
\bibliographystyle{unsrt}

\end{document}